**Title:** Strong damping-like spin-orbit torque and tunable Dzyaloshinskii-Moriya interaction generated by low-resistivity $Pd_{1-x}Pt_x$ alloys

*Lijun Zhu,\* Kemal Sobotkiewich, Xin Ma, Xiaoqin Li, Daniel. C. Ralph, Robert A. Buhrman\**

Dr. L. J. Zhu, Prof. R. A. Buhrman

Cornell University, Ithaca, NY 14850, USA

Email: lz442@cornell.edu; rab8@cornell.edu

 K. Sobotkiewich, Dr. X. Ma, Prof. X. Li

Department of Physics, Center for Complex Quantum Systems, The University of Texas at Austin, Austin, Texas 78712, USA

Prof. D. C. Ralph

Cornell University, Ithaca, NY 14850, USA

Kavli Institute at Cornell, Ithaca, New York 14853, USA

**Abstract**: Despite their great promise for providing a pathway for very efficient and fast manipulation of magnetization at the nanoscale, spin-orbit torque (SOT) operations are currently energy inefficient due to a low damping-like SOT efficiency per unit current bias, and/or the very high resistivity of the spin Hall materials. Here, we report an advantageous spin Hall material, $Pd_{1-x}Pt_x$, which combines a low resistivity with a giant spin Hall effect as evidenced through the use of three independent SOT ferromagnetic detectors. The optimal $Pd_{0.25}Pt_{0.75}$ alloy has a giant internal spin Hall ratio of >0.47 (damping-like SOT efficiency of ~ 0.26 for all three ferromagnets) and a low resistivity of ~57.5 $\mu\Omega$ cm at 4 nm thickness. Moreover, we find the Dzyaloshinskii-Moriya interaction (DMI), the key ingredient for the manipulation of chiral spin arrangements (e.g. magnetic skyrmions and chiral domain walls), is considerably strong at the $Pd_{1-x}Pt_x/Fe_{0.6}Co_{0.2}B_{0.2}$ interface when compared to that at $Ta/Fe_{0.6}Co_{0.2}B_{0.2}$ or $W/Fe_{0.6}Co_{0.2}B_{0.2}$ interfaces and can be tuned by a factor of 5 through control of the interfacial spin-orbital coupling via the heavy metal composition. This work establishes a very effective spin current generator that combines a notably high energy efficiency with a very strong and tunable DMI for advanced chiral spintronics and spin torque applications.

# 1. Introduction

Current-induced spin-orbit torques (SOTs) in heavy metal/ferromagnet (HM/FM) systems have promise for more efficient and faster electrical manipulation of magnetization at the nanoscale than magnetic field and conventional spin-transfer torque.[1-7] The damping-like SOT generated by the *bulk* spin Hall effect (SHE) of the HMs is of particular interest in exciting magnetization dynamics at microwave and terahertz frequencies,[1,2] driving skyrmion and chiral domain wall displacement[3,4] or switching the magnetization of thin film nanomagnets.[5-7] Despite extensive efforts,[1-9] the energy efficiency of present SOT operations is still limited by a relatively low damping-like SOT efficiency ($\xi_{DL}$)[6] and/or a high resistivity ($\rho_{xx}$) of the spin Hall materials[8,9] (see Table I). When $\rho_{xx}$ of the spin Hall material is comparable to or larger than that of the metallic FM being manipulated, the detrimental shunting of applied current through the FM layer will be significant (see Supporting Information). Another important ingredient for spin-orbit coupling (SOC) phenomena is the Dzyaloshinskii-Moriya interaction (DMI) at HM/FM interfaces due to the combination of the broken inversion symmetry and *interfacial* spin-orbit coupling (SOC).[10] The interfacial DMI is a short-range antisymmetric exchange interaction that can promote chiral spin alignments, e.g. in magnetic skyrmions and chiral domain walls.[3,4] The sign and magnitude of the interfacial DMI influence the direction and velocity of a chiral spin texture movement driven by damping-like SOT. For a perpendicularly magnetized multi-domain HM/FM structure, the interfacial DMI represents an obstacle for SOT switching of magnetization via a thermally-activated reversal domain nucleation and domain wall depinning process,[5] because it requires an external field or its equivalent that is larger than the DMI field applied along the bias current direction in order to switch the magnetization. This requirement leads to additional complexity in the design and implementation of SOT devices. The DMI may also play a role in the sub-ns micromagnetic dynamics that are critical to the reliable switching of nanoscale SOT-magnetic random access memories (MRAMs).[11,12] Therefore, exploration of new, more efficient HM spin Hall materials and new routes to tune the DMI strength continues to be of significant scientific interest and technological importance.

Pt is a particularly interesting spin Hall material due to its giant intrinsic spin Hall conductivity ($\sigma_{SH}$) arising from the Berry curvature of its band structure.[13,14] However, the reported values of $\xi_{DL}$ for Pt/FM systems are generally low, e.g. ~0.07 in Pt/Ni$_{81}$Fe$_{19}$ bilayers ($\rho_{xx} \approx 20$ μΩ cm),[15] ~0.12 in Pt/Co bilayers ($\rho_{xx} \approx 30$ μΩ cm).[16] Recently, the introduction of impurities[17] or disorder[18] has been found to raise $\rho_{xx}$ and to degrade $\sigma_{SH}$ of Pt (the band structure) at the same time. A good trade-off between $\rho_{xx}$ and $\sigma_{SH}$ can enhance $\xi_{DL}$ of Pt to some degree (<0.16[18]) because $\xi_{DL}=T_{int}(2e/\hbar)\sigma_{SH}\rho_{xx}$ for the intrinsic SHE mechanism. Alloying Pt with Au was recently found to be more effective than introducing impurities and disorders in enhancing the $\xi_{DL}$ as it allows significant increase in $\rho_{xx}$ without degrading $\sigma_{SH}$ in the fcc alloy with the optimized composition.[14]

In this work, based on direct spin-torque measurements using perpendicular magnetic anisotropy (PMA) Co, in-plane magnetic anisotropy (IMA) Co, and IMA Fe$_{0.6}$Co$_{0.2}$B$_{0.2}$ as the FM detectors, we report an effective magnification of the damping-like SOT efficiency and internal spin Hall conductivity (i.e. $\xi_{DL}$ and $\sigma_{SH}$) in Pd$_{1-x}$Pt$_x$ alloy. A large $\xi_{DL}$ of ≈ 0.26 and a giant $\sigma_{SH}$ of $1.1\times10^6$ $\hbar/2e$ Ω$^{-1}$m$^{-1}$ was obtained in Pd$_{0.25}$Pt$_{0.75}$ which still has a relatively low resistivity of ~57.5 μΩ cm, making Pd$_{0.25}$Pt$_{0.75}$ a strong and particularly advantageous spin Hall material from the viewpoint of energy efficiency of spintronic applications. We also find from Brillouin light scattering (BLS) measurements that the DMI at Pd$_{1-x}$Pt$_x$/Fe$_{0.6}$Co$_{0.2}$B$_{0.2}$ interfaces is both considerably strong and variable over a wide range (5×) by controlling the interfacial SOC via the HM composition.

## 2. Results and Discussions

### 2.1. Sample details

We sputter deposited Pd$_{1-x}$Pt$_x$ $d$/Co $t$ bilayers and Pd$_{1-x}$Pt$_x$ $d$/Fe$_{0.6}$Co$_{0.2}$B$_{0.2}$ $t$ bilayers (here $t$ and $d$ are thicknesses in nm) with different Pt concentration ($x$ = 0, 0.25. 0.5, 0.75, and 1) onto oxidized Si substrates (**Figure 1a**). X-ray diffraction $\theta$-$2\theta$ patterns (**Figure S1,** Supporting Information) show that the fcc $Pd_{1-x}Pt_x$ (111) diffraction peak gradually shifts with increasing $x$, affirming the uniform alloying of the two elements. For this study, the Co layers were wedges with thickness varying between 0.75-1.4 nm for $x = 1$ and between 0.64-0.94 nm for $x \leq 0.75$ across the wafer to enable the study of both PMA and IMA $Pd_{1-x}Pt_x$/Co devices. For $Pd_{1-x}Pt_x$/Co bilayers, the saturation magnetization ($M_s$) varies between 1450 and 1700 emu/cm³ for different $x$, indicative of an enhancement due to a magnetic proximity effect at the interface,[19] which, however, should not degrade $\xi_{DL}$ as is discussed in Ref. [20,21]. $M_s$ for the $Pd_{1-x}Pt_x$/$Fe_{0.6}Co_{0.2}B_{0.2}$ bilayers remains almost constant with $x$, ~1200 emu/cm³. The stacks were patterned into 5×60 μm² Hall bars by ultraviolet photolithography and ion milling. The magnetic bilayer samples were further patterned into 5×60 μm² Hall bars for measuring SOTs by out-of-plane (PMA bilayers) and in-plane (IMA bilayers) harmonic response measurements[22,23] (see **Figure** S2 and S3 in Supporting Information for more details).

## 2.2. Tuning the SHE by composition

We determined $\rho_{xx}$ for 4 nm $Pd_{1-x}Pt_x$ layers by subtracting the sheet conductance of reference stacks Ta 1/Co $t$/MgO 2/Ta 1.5 and Ta 1/ $Fe_{0.6}Co_{0.2}B_{0.2}$ $t$/MgO 2/Ta 1.5 from that of our samples containing the $Pd_{1-x}Pt_x$ layer. As plotted in **Figure** 1b, $\rho_{xx}$ for 4 nm $Pd_{1-x}Pt_x$ layers varies between 37.0 and 57.6 μΩ cm for different $x$. The alloys with $0.25 \leq x \leq 0.75$ have greater $\rho_{xx}$ than pure Pt or Pd for all three FM cases, which we attribute to enhanced electron scattering in the chemically disordered alloys. At a given $x$ for the different FM samples, there are only small differences in $\rho_{xx}$ likely due to modest differences in interfacial scattering. **Figure** 1c shows the $x$ dependence of $\xi_{DL}$ generated by 4 nm $Pd_{1-x}Pt_x$ for PMA Co ($t = 0.64$ or 0.75), IMA Co ($t = 0.94$ or 1.4), and IMA $Fe_{0.6}Co_{0.2}B_{0.2}$ ($t = 2.8$) detectors. Here $\xi_{DL} = 2e\mu_0 M_s t H_{DL}/\hbar j_e$, where $\mu_0$, $M_s$, and $H_{DL}$ are the permeability of vacuum, saturation magnetization, and the damping-like SOT effective field. For all three FM cases, $\xi_{DL}$ increases quickly from ~0.07 at $x = 0$ (pure Pd) to the peak value of ~0.26 at $x = 0.75$, and then drops to ~0.20 at $x = 1$ (pure Pt). The consistent peak behavior of $\xi_{DL}$ at $x \approx 0.75$ is attributable to the non-monotonic $x$ dependence of $\rho_{xx}$ (Figure 1b) and the giant apparent spin Hall conductivity $\sigma_{SH}^* = T_{int}\sigma_{SH} = (\hbar/2e)\xi_{DL}/\rho_{xx}$ for $x \geq 0.75$. As seen in Figure 1d, $\sigma_{SH}^*$ for 4 nm $Pd_{1-x}Pt_x$ increases monotonically with $x$ and exhibits a weak peak at $x \approx 0.75$ for the PMA case. Interestingly, the $x$ dependence is *functionally* similar to the predicted behavior for the intrinsic spin Hall conductivities of $Pd_{1-x}Pt_x$ alloys in a recent *ab initio* calculation,[14] although as we will discuss below the actual spin Hall conductivities for $Pd_{1-x}Pt_x$ are much larger than indicated by the *ab initio* results in Ref. [14] once the degradation of $T_{int}$ by spin backflow (SBF) and spin memory loss (SML) at the FM interfaces[24] is taken into account. Finally, we point out that for each FM case, $\sigma_{SH}^*$ of the 4 nm $Pd_{0.25}Pt_{0.75}$ is comparable with that of pure Pt, i.e. $(4.6\text{-}5.1)\times10^5$ $\hbar/2e$ $\Omega^{-1}$ $m^{-1}$. This is distinct from the case of introduing Hf impurites into Pt,[17] where $\sigma_{SH}^*$ is degraded by > 25% after the Hf impurity concentration reaches 12.5%. Another interesting observation is that $\sigma_{SH}^*$ for each $x$ is overall *slightly* different for the three FM detectors, i.e. $\sigma_{SH}^*$ ($Fe_{0.6}Co_{0.2}B_{0.2}$ IMA) > $\sigma_{SH}^*$ (Co IMA) > $\sigma_{SH}^*$ (Co PMA). This difference is attributable in part to the Co thickness in the PMA sample being less than its spin dephasing length (~1 nm), because of which the spin current is not completely absorbed after going through the FM layer,[16] and in part, we speculate, to differences in SBF and SML at the different interfaces. The maximum $\xi_{DL} \approx 0.26$ for 4 nm $Pd_{0.25}Pt_{0.75}$ is comparable to that of highly resistive $\beta$-W (≈300 μΩ cm),[8] while $\rho_{xx}$ is as low as ~57.5 μΩ cm, less than one-fourth of that of $\beta$-W with similar thickness.[8] It is interesting to notice that $\xi_{DL}$ for 4 nm $Pd_{0.25}Pt_{0.75}$ is even slightly smaller than ~0.30 for 4 nm $Au_{0.25}Pt_{0.75}$ (≈ 80 μΩ cm)[14] due to its low $\rho_{xx}$ despite that their similar $\sigma_{SH}^*$. We note also that $\xi_{DL} \approx 0.2$ for the PMA Pt 4/Co 0.75 bilayers ($\rho_{xx}$ ≈51 μΩ cm, see Figure 1b) is larger than ~0.12 (~0.08) as previously reported for Pt/Co (Pt/$Ni_{81}Fe_{19}$) bilayers where $\rho_{xx}$ was lower, as small as 30 (20) μΩ cm.[14,16] This difference in $\rho_x$ and hence in $\xi_{DL}$ is most likely due to differences in film growth protocols. Finally our result of $\xi_{DL} = 0.07$ for Pd is comparable with that reported for Pd/Co/$AlO_x$ (~0.06).[25]

Now we consider possible reasons for the absence of the strong SHE and enhanced damping-like SOT efficiency in two previous reports.[26,27] Zhou *et al.*[26] reported from a spin-torque ferromagnetic resonance (ST-FMR) study that the spin Hall angle ($\theta_{SH}$) of $Pd_{1-x}Pt_x$ was enhanced to ~0.05 in the composition range $0.5 \leq x \leq 0.8$ compared to -0.005 for Pd and 0.04 for Pt. In another inverse SHE/spin pumping work,[27] $\theta_{SH}$ of $Pd_{1-x}Pt_x$ was found to be ~0.17 at $x\approx0.8$, which is larger than 0.045 in Pd to 0.125 in Pt. Despite the similar trends as we observe, the two experiments seem to considerably underestimate the SHE strength of the alloys. We note that the ST-FMR analysis ignored any influence of SBF, SML, field-like SOT, and the measurement artifacts (e.g. a strong dependence of the SOT efficiency on the microwave frequency). Meanwhile, both spin diffusion length ($\lambda_s$) and $\rho_{xx}$ of the HMs, which are crucial for the analysis of an inverse SHE experiment, were assumed to be independent of the HM thickness, which is however an invalid assumption as discussed below and in Ref. [28].

## 2.3. Spin backflow and internal spin Hall effect

The spin diffusion length ($\lambda_s$) and spin conductance $1/\lambda_s\rho_{xx}$ are two key parameters for the theoretical understanding of a spin Hall material, for determining the spin Hall ratio ($\theta_{SH}$) of a HM via an inverse SHE experiment, and for optimizing the SOT effectiveness of the material. We determined $1/\lambda_s\rho_{xx}$ for $Pd_{0.25}Pt_{0.75}$, the optimal composition for the largest $\xi_{DL}$ (**Figure 1c**) by studying $H_{DL}$ and $\xi_{DL}$ for a series of $Pd_{0.25}Pt_{0.75}$ $d$/Co 0.64 bilayers as a function of $d$. As plotted in **Figure 2a,** $H_{DL}/E$ initially increases quickly and then gradually saturates with increasing $d$, consistent with the spin diffusion behavior expected by the SHE model. $\xi_{DL}$, however, first increases as $d$ increases from 2 nm, peaks at $d$ =3-4 nm, and then drops gradually as $d$ further increases (**Figure 2b).** This is a direct consequence of the combination of the decreasing $\rho_{xx}$ with $d$ as interfacial scattering becomes less dominant (**Figure 2c**) and the rapid and then saturating increase in $\sigma_{SH}^*$ (the black squares in **Figure 2d**) as $d$ increases up to and then beyond the HM $\lambda_s$. Overall, the average resistivity $\rho_{xx}$ of the $Pd_{0.25}Pt_{0.75}$ layers drops from 86.8 μΩ cm at $d$ = 2 nm to 37.4 μΩ cm at $d$ = 8 nm. Such an interesting peak behavior of $\xi_{DL}$ and the strong thickness dependence of $\rho_{xx}$ were not observed in $Au_{0.25}Pt_{0.75}$ where the mean-free path was very short. [14]

If we assume the usual case where the interfacial spin mixing conductance $G^{\uparrow\downarrow} \approx \mathrm{Re}\,G^{\uparrow\downarrow}$, a negligible SML at the HM/FM interface, and a $\rho_{xx}$-independent $\lambda_s$, both $\lambda_s$ and $\sigma_{SH}$ can be obtained by [28]

$$\sigma_{SH}^{*} = \sigma_{SH}(1 - \mathrm{sech}(d/\lambda_s))(1 + \tanh(d/\lambda_s)/2\lambda_s\rho_{xx}\mathrm{Re}\,G^{\uparrow\downarrow})^{-1} \qquad [1]$$

However, this equation is not valid for $Pd_{0.25}Pt_{0.75}$ as $\rho_{xx}$ varies strongly with $d$ and $\lambda_s$ is not a constant if the Elliot-Yafet spin scattering mechanism dominates, where $\lambda_s \propto 1/\rho_{xx}$. Considering this effect, we can use the "rescaling" method introduced in Ref. [28] to analyze our data. In Figure 2d, we plot $\sigma_{SH}^{*}$ as a function of the rescaled thickness $d_0$ (the red circles) which uses 57.5 $\mu\Omega$ cm, the average resistivity of the 4 nm $Pd_{0.25}Pt_{0.75}$ film, as the constant reference resistivity ($\rho_{xx0}$). Using $G^{\uparrow\downarrow} \approx 6\times10^{14}$ $\Omega^{-1}\mathrm{m}^{-2}$ as an approximation as calculated for Pt/Co interfaces,[29] the best fit of $\sigma_{SH}^{*}$ vs $d_0$ to the above Equation gives $\sigma_{SH} \approx (1.05 \pm 0.02)\times10^{6}$ $\hbar/2e$ $\Omega^{-1}$ $\mathrm{m}^{-1}$, $\lambda_s \approx 1.95\pm0.12$ nm (for 4 nm thickness), and $T_{int}$ of 0.55. Accordingly, the internal $\theta_{SH}$ for 4 nm $Pd_{0.25}Pt_{0.75}$ is determined to be $> 0.47$ after the SBF correction. From these results we find $1/\lambda_s\rho_{xx}$ for $Pd_{0.25}Pt_{0.75}$ to be $\approx 0.89\times10^{15}$ $\Omega^{-1}\mathrm{m}^{-2}$, which is slightly larger than $0.71 \times10^{15}$ $\Omega^{-1}\mathrm{m}^{-2}$ for $Au_{0.25}Pt_{0.75}$ but considerably lower than $1.3\times10^{15}$ $\Omega^{-1}\mathrm{m}^{-2}$ obtained for Pt from a similar Pt thickness dependent study.[28] The relatively low spin conductance of $Pd_{0.25}Pt_{0.75}$ is advantageous for reducing SBF at $Pd_{0.25}Pt_{0.75}$ /FM interfaces. We also note that the large value for $\sigma_{SH}$ is perhaps still a lower bound as it assumes an ideal $Pd_{1-x}Pt_x$/Co interface and does not take into account any SML induced by interfacial spin-orbit scattering. This lower-bound spin Hall conductivity of $\sim1.1\times10^{6}$ $\hbar/2e$ $\Omega^{-1}$ $\mathrm{m}^{-1}$ is substentially larger than $5.9\times10^{5}$ $\hbar/2e$ $\Omega^{-1}$ $\mathrm{m}^{-1}$ of pure Pt [28] and $7.0\times10^{5}$ $\hbar/2e$ $\Omega^{-1}$ $\mathrm{m}^{-1}$ for the $Au_{0.25}Pt_{0.75}$ alloy[14], which indicates a significant enhancement in the spin Hall conductivity that warrant future theoretical efforts.

## 2.4. Strong and tunable interfacial DMI

Since the interfacial DMI strength is an important factor in many SOT phenomena and applications,[3-5] we quantitatively determined the interfacial DMI in $Pd_{1-x}Pt_x$ 4/$Fe_{0.6}Co_{0.2}B_{0.2}$ 2.6 bilayers by measuring the DMI-induced frequency difference ($\Delta f_{DM}$) between counter-propagating Damon-Eshbach spin waves using BLS. Figure 3a shows the BLS measurement geometry, where a magnetic field $H$ = 1700 Oe was applied along the $x$ direction to align the magnetization of the FM layer. The total in-plane momentum is conserved during the scattering process. The anti-Stokes (Stokes) peaks in BLS spectra correspond to the annihilation (creation) of magnons with a wave-vector $k = \pm 4\pi\sin\theta/\lambda$, where $\theta$ is the light incident angle w.r.t. the film normal, and $\lambda$ = 532 nm is the laser wavelength. In Figure 3b we plot $\Delta f_{DM}$ as a function of $k$ for different composition $x$. Here $\Delta f_{DM}$ is the frequency difference of the $\pm k$ peaks and averaged for $\pm H$ (see more details in ref. [30]), as can be seen from the representative BLS spectrum in Figure 3c ($|k|$ = 9.6 $\mu m^{-1}$, $H$ = 1700 Oe, $x$ = 1). The linear relation between $\Delta f_{DM}$ and $k$ for each $x$ agrees with the expected relation $\Delta f_{DM} \approx (2\gamma/\pi\mu_0 M_s)Dk$,[31,32] where $\gamma \approx$ 176 GHz/T is the gyromagnetic ratio and $D$ is the volume-averaged (volumetric) DMI constant over the FM thickness. As shown in Figure 3d, with increasing $x$, $D$ was tuned by a factor of ~5, i.e. from -0.57 to -0.12 erg/cm$^2$. This strong tunability of interfacial DMI is mainly attributed to the variation of the SOC strength at the $Pd_{1-x}Pt_x$/ $Fe_{0.6}Co_{0.2}B_{0.2}$ interface as indicated by the linear relation between $D$ and the interfacial magnetic anisotropy energy density ($K_s$) as determined from $Fe_{0.6}Co_{0.2}B_{0.2}$ thickness dependent anisotropy studies using spin-torque ferromagnetic resonance (see Figure 3e and Figure S4). We interpret the intercept of the linear $K_s$-$D$ fit as representing the contribution of the $Fe_{0.6}Co_{0.2}B_{0.2}$/MgO interface to the overall interfacial anisotropy energy density, i.e $K_s$ (MgO) ace to the ov$^2$ for these as-grown/unannealed samples, while $K_s$-$K_s$ (MgO) is the contribution from the $Pd_{1-x}Pt_x$/$Fe_{0.6}Co_{0.2}B_{0.2}$ interface and is an indicator of the interfacial SOC strength. Taking into account the inverse dependence of $D$ on FM thickness $t$ due to the volume averaging effect during BLS measurement, the total DMI strength for HM/$Fe_{0.6}Co_{0.2}B_{0.2}$ interface can be estimated as $D_s = Dt$. For the $Pd_{1-x}Pt_x$/$Fe_{0.6}Co_{0.2}B_{0.2}$ interfaces, $D_s$ changes from -1.47×10$^{-7}$ erg/cm at $x$ = 1 to -0.30×10$^{-7}$ erg/cm at $x$ = 0. We note first that the DMI for our $Pd_{1-x}Pt_x$/$Fe_{0.6}Co_{0.2}B_{0.2}$ interface is very strong compared with those for Ta (or W) / $Fe_{0.6}Co_{0.2}B_{0.2}$ systems,[30] i.e. $D_s$ (Ta) = 0.36×10$^{-8}$ erg/cm and $D_s$ (W) = 0.73×10$^{-8}$ erg/cm. Secondly, the $Pd_{1-x}Pt_x$ composition enables a rather significant tunability of $D_s$ because it is equivalent to a change from 8.3 (4.1) to 40.8 (20.1) times that of Ta (W)/$Fe_{0.6}Co_{0.2}B_{0.2}$ systems.[32] The large amplitude of the DMI and its tunability ($\geq 3\times$) for $x \geq 0.5$ where $\xi_{DL} \geq 0.2$, make $Pd_{1-x}Pt_x$ especially intriguing for developing new chiral spintronic devices and for exploring new DMI effects on the performance of skyrmion, chiral domain wall devices, and micromagnetics during SOT magnetization switching. The quite different composition dependence of the interfacial DMI strength and damping-like SOT further confirms that the physical source of the observed strong SOTs in the $Pd_{1-x}Pt_x$/FM systems is dominated by the *bulk* SHE other than the *interfacial* SOC. Note that while the interfacial DMI is mainly dertermined by the *interfacial* SOC that is sensitive to the short-range ordering at the interface,[10] the intrinsic bulk SHE is determined by the *bulk* SOC-rated topology of the band structure of the HMs. [13]

## 2.5. High energy efficiency in spin-torque applications

We finally emphasize that the low $\rho_{xx}$ and giant $\xi_{DL}$ make $Pd_{0.25}Pt_{0.75}$ advantageous for SOT research and technological applications with metallic FMs. As a simple example, we first show in Figure 4a that damping-like SOT generated by the SHE of a 4 nm $Pd_{0.25}Pt_{0.75}$ can switch the magnetization of a Co 0.64 layer (with an effective PMA field of 7.7 kOe and coercivity of 0.44 kOe) with a dc current of ~4.8 mA (corresponding to $j_e = 2.2\times10^7$ A/cm$^2$ in the $Pd_{0.25}Pt_{0.75}$ layer) and a bias field $H_x = \pm100$ Oe applied along the current direction. This reaffirms the giant $\xi_{DL}$ of $Pd_{0.25}Pt_{0.75}$. To better illustrate the advantage of $Pd_{0.25}Pt_{0.75}$, in Table I we compare the parameters that are most important for SOT applications, i.e. $\xi_{DL}$, $\rho_{xx}$, and $\sigma_{SH}^*$, for the various strong spin Hall materials. In Figure 4b and Table 1, we further compare the calculated write power consumption ($P$) for a typical in-plane magnetized SOT-MRAM device, which curently is of the most technological promise for very fast (~200 ps), low-current density, low-error rate,[7] and field-free switching,[33] based on these spin Hall materials by taking into consideration current shunting into the FeCoB free layer (see Supporting Information). Here we used a 600×300×4 nm$^2$ spin Hall channel, a 190×30×1.8 nm$^3$ FeCoB free layer (resistivity ≈130 μΩ cm) and the parallel resistor model for the illustrative calculation (Supporting Information).

The power consumption for this modeled $Pd_{0.25}Pt_{0.75}$-based MRAM device is 320%, 50% and 20% smaller than low-resistivity Pt (20 μΩ cm)[15], high-resistivity Pt (51 μΩ cm), and $Au_{0.25}Pt_{0.75}$. We note that $\xi_{DL}$ of 0.2 for the high-resistivity pure Pt is the highest among the reported Pt values. The $Pd_{0.25}Pt_{0.75}$-based MRAM device is 8 and 23 times smaller than those based on $\beta$-W (300 μΩ cm),[8] and $\beta$-Ta (190 μΩ cm),[6] respectively. An MRAM device based on a magnetic tunnel junction with metallic ferromagnetic electrodes in combination with a $Pd_{0.25}Pt_{0.75}$ spin Hall channel is estimated to be markedly more efficient in energy than those that might utilize the topological insulators $Bi_2Se_3$ ($\xi_{DL}$=3.5),[34] $(Bi,Se)_2Te_3$ ($\xi_{DL}$=0.4),[9] and $Bi_xSe_{1-x}$ ($\xi_{DL}$=18.6),[35] mainly because the colossal resistivity of the topological insulators results in very strong current shunting into the free layer and giant energy consumption in the channel. The small resistivity also makes $Pd_{0.25}Pt_{0.75}$ more compelling than $Au_{0.25}Pt_{0.75}$ [14] for certain device applications, e.g. cryogenic computing, where the transistor circuits require a small write impedance for magnetic memory devices.[36] Moreover, $Pd_{0.25}Pt_{0.75}$ is compatible with both sputtering techniques and the use of Si substrates, which are preferable for integration technology. Therefore, the combination of the giant $\xi_{DL}$, the low $\rho_{xx}$, and the compatibility with microelectronics manufacturing technology makes $Pd_{0.25}Pt_{0.75}$ a particularly attractive spin Hall material for the generation and detection of spin currents in spintronic devices.

## 3. Conclusions

We have established via direct SOT measurements a strong spin Hall material $Pd_{0.25}Pt_{0.75}$ which has a giant internal spin Hall ratio of > 0.47 (yielding $\xi_{DL} \approx 0.26$ for its bilayers with either Co or $Fe_{0.6}Co_{0.2}B_{0.2}$), spin Hall conductivity of > $1.1 \times 10^6$ $\hbar/2e$ $\Omega^{-1}m^{-1}$, spin conductance $1/\lambda_s\rho_{xx}$ of $0.89 \times 10^{15}$ $\Omega^{-1}m^{-2}$, and a relatively low $\rho_{xx}$ of ~57.5 μΩ cm (at 4 nm thickness). Particularly, this giant SHE and low $\rho_{xx}$ make $Pd_{0.25}Pt_{0.75}$ more energy-efficient for manipulating metallic magnetic devices than the other HMs (W, Ta, Pt, and $Au_{0.25}Pt_{0.75}$) and the topological insulators $Bi_xSe_{1-x}$, $Bi_2Se_3$ and $(Bi,Se)_2Te_3$. We also find the interfacial DMI at $Pd_{1-x}Pt_x/Fe_{0.6}Co_{0.2}B_{0.2}$ interfaces is very strong and also widely tunable (×5) by Pt concentration $x$; especially for $x \geq 0.5$ where $\xi_{DL} \geq 0.2$, the interfacial DMI can be tuned by a factor of 3. Our findings provide a highly efficient spin Hall material system that simultaneously combines a giant SHE, low resistivity, strong and tunable DMI, with excellent processing compatibility for device integration, for developing new efficient SOT-driven magnetic memories and chiral (skyrmion and chiral domain walls) spintronic devices.

## 4. Experimental Section

***Sample growth and characterizations.*** The magnetic bilayers of $Pd_{1-x}Pt_x$ $d$/Co or $Fe_{0.6}Co_{0.2}B_{0.2}$ $t$ with different Pt concentration ($x = 0, 0.25. 0.5, 0.75$, and 1) were sputter deposited at room temperature on oxidized Si substrates and capped with MgO 2/Ta 1.5 protective layers where the Ta layer was fully oxidized upon exposure to the atmosphere. A seed layer of Ta 1 was grown prior to the co-sputtering of $Pd_{1-x}Pt_x$ layer in order to improve the adhesion and smoothness of the magnetic bilayers. The saturation magnetization of each sample was measured by a standard vibrating sample magnetometer embedded in a Quantum Design physical properties measurement system. A Rigaku Smartlab x-ray diffractometer was introduced to characterization the structure.

***Device fabrication and measurement.*** The stacks were patterned into $5 \times 60$ $\mu m^2$ Hall bars by photolithography and ion milling for harmonic response measurements. A lock-in amplifier is introduced to source a 4 V sinusoidal voltage onto the Hall bar along the $x$ axis (corresponding to an alternating electrical field of constant magnitude $E \approx 66.7$ kV/m) and to detect the in-phase first and out-of-phase second harmonic Hall voltages. The current switching measurements were performed with a Yokogawa 76541 current source and a Keithley 2000 voltage meter.

## Supporting Information

Supporting Information is available from the Wiley Online Library or from the author.

## Acknowledgements

This work was supported in part by the Office of Naval Research, by the NSF MRSEC program (DMR-1719875) through the Cornell Center for Materials Research, and by the Office of the Director of National Intelligence (ODNI), Intelligence Advanced Research Projects Activity (IARPA), via contract W911NF-14-C0089. The views and conclusions contained herein are those of the authors and should not be interpreted as necessarily representing the official policies or endorsements, either expressed or implied, of the ODNI, IARPA, or the U.S. Government. The U.S. Government is authorized to reproduce and distribute reprints for Governmental purposes notwithstanding any copyright annotation thereon. The DMI measurement performed at UT-Austin was supported by SHINES, an Energy Frontier Research Center funded by the U.S. Department of Energy (DoE), Office of Science, Basic Energy Science (BES) under award # DE-SC0012670. The devices were fabricated in part at the Cornell NanoScale Facility, an NNCI member supported by NSF Grant ECCS-1542081.

**Conflict of Interest**

The authors declare no conflict of interest.

**Key words**: spin-orbit torque, spin Hall conductivity, spin Hall effect, Dzyaloshinskii-Moriya interaction

--------------------------------------------------------------------------

**References**

[1] V. E. Demidov, S. Urazhdin, H. Ulrichs, V. Tiberkevich, A. Slavin, D. Baither, G. Schmitz, S. O. Demokritov, *Nat. Mater*. **2012,** 11, 1028.

[2] Y. Wu, M. Elyasi, X. Qiu, M. Chen, Y. Liu, L. Ke, H. Yang, *Adv. Mater*. **2017**, 29, 1603031.

[3] W. Jiang, P. Upadhyaya, W. Zhang, G. Yu, M. B. Jungfleisch, F. Y. Fradin, J. E. Pearson, Y.Tserkovnyak, K. L. Wang, O. Heinonen, S. G. E. Velthuis, A. Hoffmann, *Science* **2015**, 349, 283-286.

[4] P. P. J. Haazen, E. Muré, J. H. Franken, R. Lavrijsen, H. J. M. Swagten,  B. Koopmans, *Nat. Mater.* **2013**, 12, 299.

[5] O. J. Lee, L. Q. Liu, C. F. Pai, Y. Li, H. W. Tseng, P. G. Gowtham, J. P. Park, D. C. Ralph,  R. A. Buhrman, *Phys. Rev. B* **2014**, 89, 024418.

[6] L. Liu, C.-F. Pai, Y. Li, H. W. Tseng, D. C. Ralph,  R. A. Buhrman, *Science*, **2012**, 336, 555.

[7] S. Shi, Y. Ou, S.V. Aradhya, D. C. Ralph,  R. A. Buhrman, *Phys. Rev. Applied* **2018**, 9, 011002.

[8] C.-F. Pai, L. Liu, Y. Li, H. W. Tseng, D. C. Ralph,  R. A. Buhrman, *Appl. Phys. Lett.* **2012**, *101*, 122404.

[9] J. Han, A. Richardella, S. A. Siddiqui, J. Finley, N. Samarth, and L. Liu, *Phys. Rev. Lett.* **2017**, *119*, 077702.

[10] H. Yang, A. Thiaville, S. Rohart, A. Fert, and M. Chshiev, *Phys. Rev. Lett.* **2015**,*115*, 267210.

[11] G. E. Rowlands, S. V. Aradhya, S. Shi, E. H. Yandel, J. Oh, D. C. Ralph, R. A. Buhrman, *Appl. Phys. Lett.* **2017**, 110, 122402.

[12] M. Cubukcu, J. Sampaio, K. Bouzehouane, D. Apalkov, A. V. Khvalkovskiy, V. Cros, N. Reyren, *Phys. Rev. B* **2016**, 93, 020401(R).

[13] E. Sagasta, Y. Omori, M. Isasa, M. Gradhand, L. E. Hueso, Y. Niimi, Y. Otani, F. Casanova, *Phys. Rev. B* **2016**, 94, 060412(R).

[14] L. Zhu, D. C. Ralph, and R. A. Buhrman, *Phys. Rev. Applied* **2018**, 10, 031001.

[15] L. Liu, T Moriyama, D. C Ralph, R. A. Buhrman, *Phys. Rev. Lett.* **2011**, 106, 036601.

[16] C.-F. Pai, Y. Ou, L. H. Vilela-Leao, D. C. Ralph, R. A. Buhrman, *Phys. Rev. B* **2015**, 92, 064426.

[17] M.-H. Nguyen, M. Zhao, D.C. Ralph, R. A. Buhrman, *Appl. Phys. Lett.* **2016**, 108, 242407.

[18] J. W. Lee, Y.-W. Oh, S.-Y. Park, A. I. Figueroa, G. van der Laan, G. Go, K.-J. Lee, B.-G. Park, *Phys. Rev. B* **2017**, 96, 064405.

[19] C. L. Canedy, X.W. Li, G. Xiao, *Phys. Rev. B* **2000**, 62, 508-519.

[20] T. A. Peterson, A. P. McFadden, C. J. Palmstrøm, P. A. Crowell, *Phys. Rev. B* **2018**, 97, 020403(R).

[21] L. Zhu, D. C. Ralph, and R. A. Buhrman, *Phys. Rev. B* **2018**, 98, 134406.

[22] M. Hayashi, J. Kim, M. Yamanouchi, H. Ohno, *Phys. Rev. B* **2014**, 89, 144425.

[23] C. O. Avci, K. Garello, M. Gabureac, A. Ghosh, A. Fuhrer, S. F. Alvarado, P. Gambardella, *Phys. Rev. B* **2014**, 90, 224427.

[24] J.-C. Rojas-Sánchez, N. Reyren, P. Laczkowski, W. Savero, J.-P. Attané, C. Deranlot, M. Jamet, J.-M. George, L. Vila, H. Jaffrès, *Phys. Rev. Lett.* **2014**, 112, 106602.

[25] A. Ghosh, K. Garello, C. O. Avci, M. Gabureac, P. Gambardella,  *Phys. Rev. Appl.* **2017**, 7, 014004.

[26] X. Zhou, M. Tang, X. L. Fan, X. P. Qiu, and S. M. Zhou, *Phys. Rev. B* **2016**, 94, 144427.

[27] L. Ma, H.-A. Zhou, L. Wang, X.-L. Fan, W.-J. Fan, D.S. Xue, K. Xia, Z. Wang, R.-Q. Wu, G.-Y. Guo, L. Sun, X. Wang, X.-M. Cheng, S.-M. Zhou, *Adv. Electron. Mater.* **2016**, 2, 1600112.

[28] M.-H. Nguyen, D. C. Ralph, R. A. Buhrman, *Phys. Rev. Lett.* **2016**, 116, 126601.

[29] P. M. Haney, H.-W. Lee, K.-J. Lee, A. Manchon, M. D. Stiles, *Phys. Rev. B* **2013**, 87, 174411.

[30] X. Ma, G. Yu, C. Tang, X. Li, C. He, J. Shi, K. L. Wang, X. Li, *Phys. Rev. Lett.* **2018**, 120, 157204.

[31] J.-H. Moon, S.-M. Seo, K.-J. Lee, K.-W. Kim, J. Ryu, H.-W. Lee, R. D. McMichael, M. D. Stiles, *Phys. Rev. B* **2013**, 88, 184404.

[32] H. T. Nembach, J. M. Shaw, M. Weiler, E. Jué, T. J. Silva, *Nat. Phys.* **2015**, 11, 825.

[33] S. Fukami, T. Anekawa, C. Zhang, H. Ohno, *Nat. Nanotech.* **2016**, 11, 621–625.

[34] A. R. Mellnik, J. S. Lee, A. Richardella, J. L. Grab, P. J. Mintun, M. H. Fischer, A. Vaezi, A. Manchon, E. −A. Kim, N. Samarth, D. C. Ralph, *Nature* **2014**, 511, 499-451.

[35] M. DC, R. Grassi, J.-Y. Chen, M. Jamali, D. R. Hickey,D. Zhang, Z. Zhao, H. Li, P. Quarterman, Y. Lv, M. Li, A. Manchon, K. A. Mkhoyan, T. Low, J.-P. Wang, *Nat. Mater.* **2018**, DOI: 0.1038/s41563-018-0136-z.

[36] D.S. Holmes, A. L. Ripple, M. A. Manheiner, IEEE Trans. Appl. Supercond. **2013**, 23, 1701610.

Table I. Comparison of $\xi_{DL}$, $\rho_{xx}$, $\sigma_{SH}^*$, and normalized power consumption $P$ for various strong spin current generators (thickness = 4 nm).

| | $\xi_{DL}$ | $\rho_{xx}$ (μΩ cm) | $\sigma_{SH}^*$ ($10^5$ ℏ/2e $\Omega^{-1}m^{-1}$) | $P$ | Ref. |
|---|---|---|---|---|---|
| $Pd_{0.25}Pt_{0.75}$ | 0.26 | 57.5 | 4.5 | 0.68 | This work |
| $Au_{0.25}Pt_{0.75}$ | 0.3 | 83 | 3.6 | 0.82 | Zhu $et\ al$ [14] |
| Pt (high-$\rho_{xx}$) | 0.20 | 51 | 3.9 | 1 | This work |
| $Bi_2Se_3$ | 3.5 | 1755 | 2.0 | 2.1 | Mellnik $et\ al$ [34] |
| Pt (low-$\rho_{xx}$) | 0.07 | 20 | 3.5 | 2.8 | Liu $et\ al$ [15] |
| $\beta$-W | 0.3 | 300 | 1.0 | 5.7 | Pai $et\ al$ [8] |
| $\beta$-Ta | 0.12 | 190 | 0.63 | 16.7 | Liu $et\ al$ [6] |
| $Bi_xSe_{1-x}$ | 18.6 | 13000 | 1.4 | 20 | DC $et\ al$ [35] |
| $(Bi,Se)_2Te_3$ | 0.4 | 4020 | 0.1 | 1474 | Han $et\ al$ [9] |

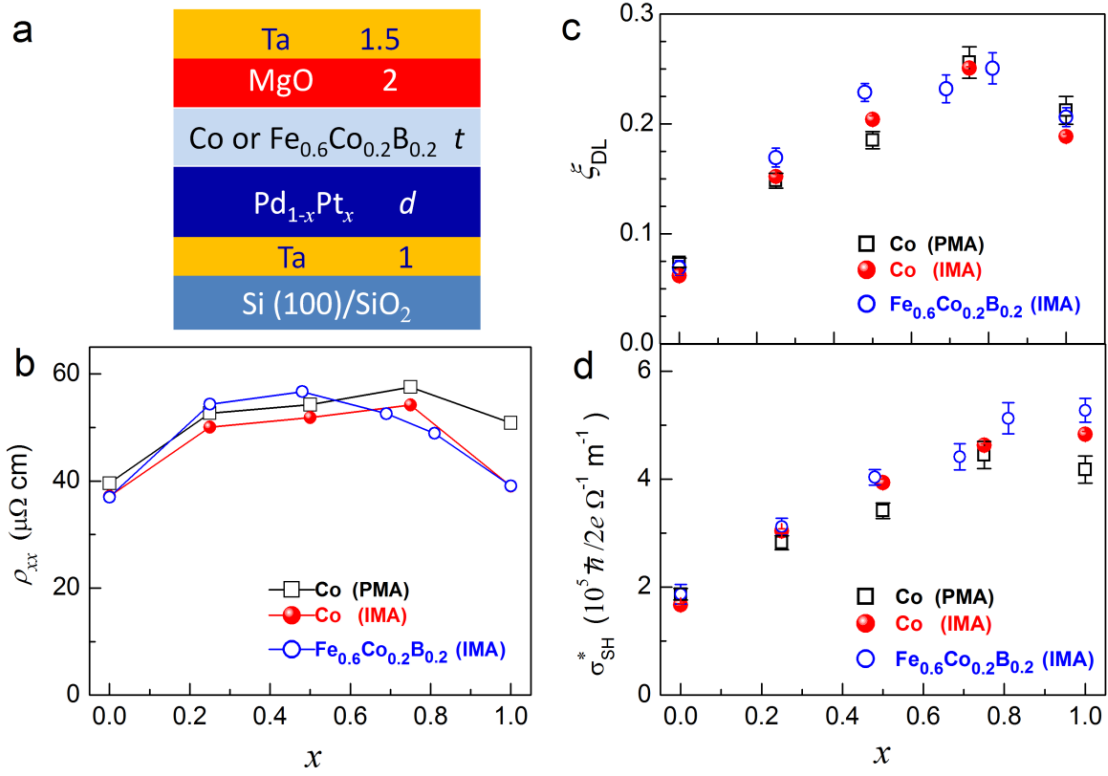

**Figure 1.** a) Schematic depiction of the magnetic stacks. $x$ dependence of b) $Pd_{1-x}Pt_x$ resistivity ($\rho_{xx}$), c) $\xi_{DL}$ and $\sigma_{SH}^*$ for 4 nm $Pd_{1-x}Pt_x$ using Co and $Fe_{0.6}Co_{0.2}B_{0.2}$ as FM detectors in the SOT measurements.

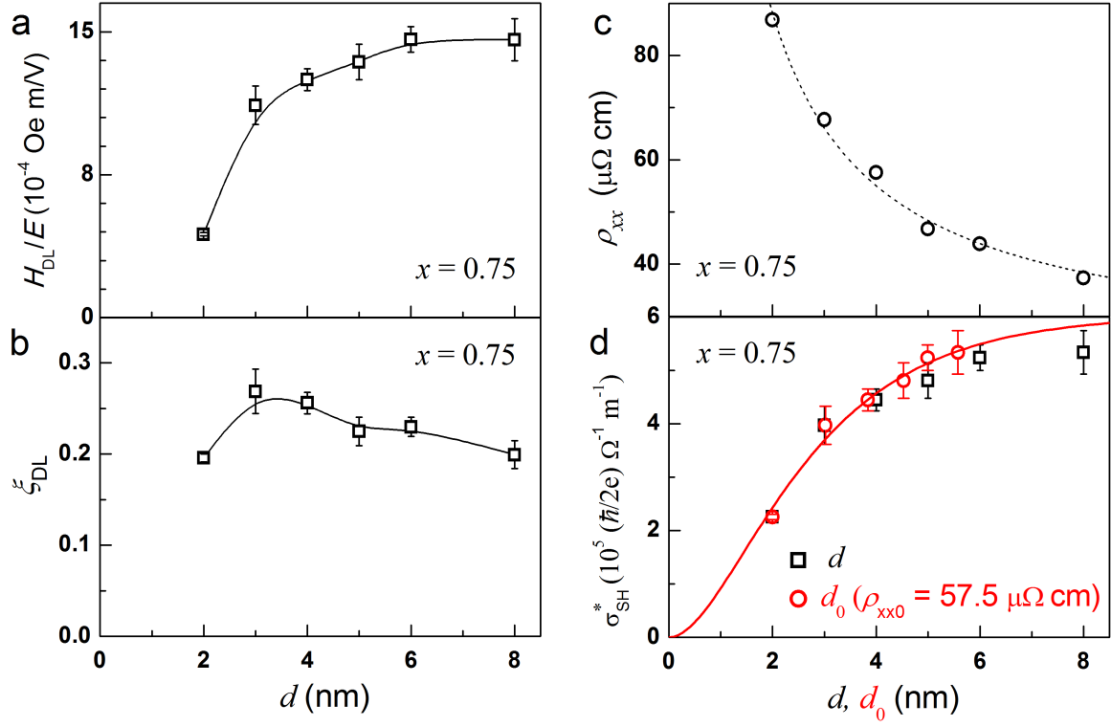

**Figure 2.** Pd$_{0.25}$Pt$_{0.75}$ thickness ($d$) dependence of a) $H_{DL}/E$, b) $\xi_{DL}$, and c) $\rho_{xx}$; d) $\sigma_{SH}^*$ plotted as a function of $d$ (black squares) and the rescaled effective thickness ($d_0$, red circles). The solid curve donates the best $\sigma_{SH}^*$-$d_0$ fit. In d), $d_0$ was rescaled with the resistivity of the 4 nm Pd$_{0.25}$Pt$_{0.75}$ film ($\rho_{xx0} = 57.5$ μΩ cm).

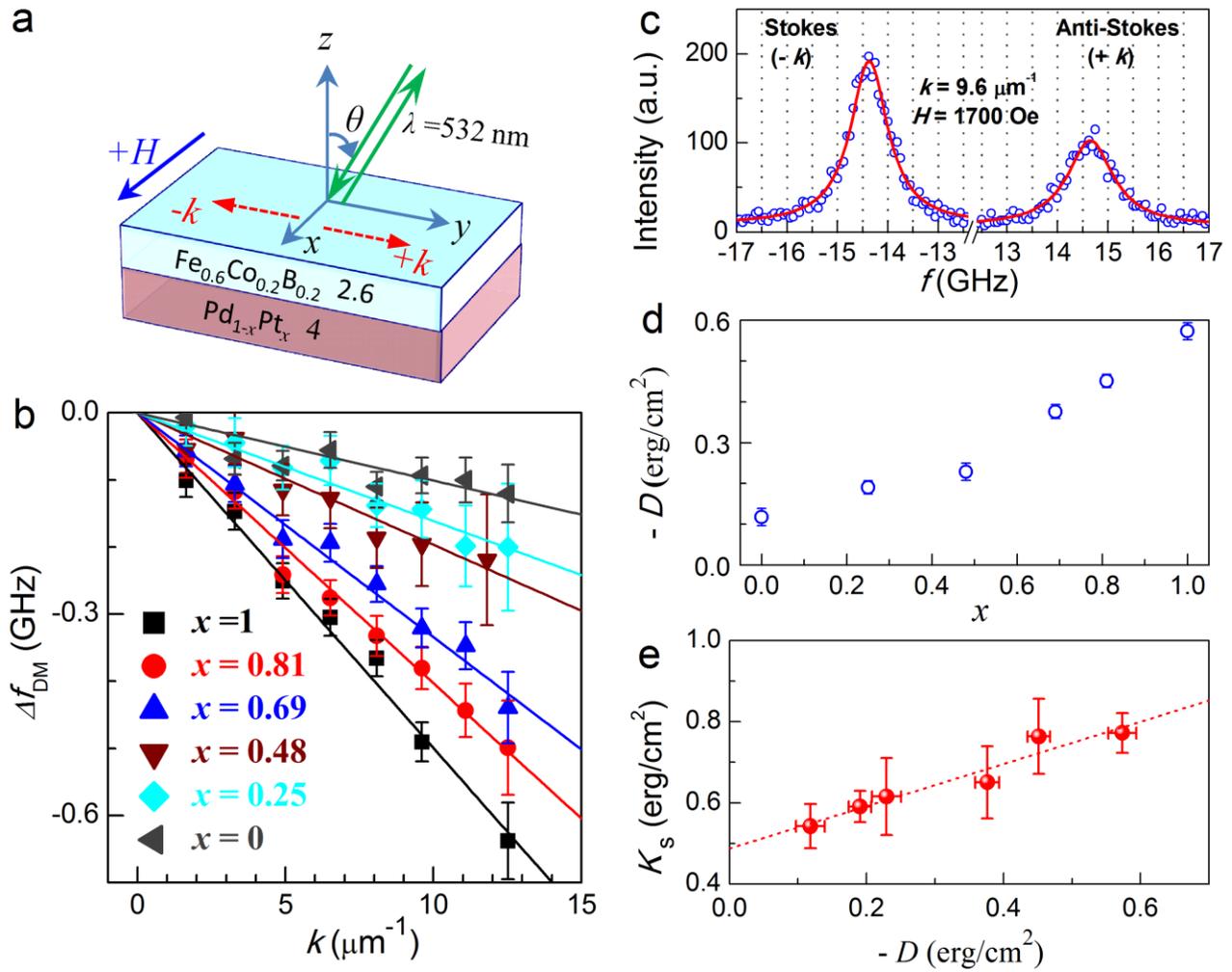

**Figure 3**. a) BLS measurement geometry; b) $k$ dependence of $\Delta f_{DM}$; c) BLS spectra at $k = 9.6\ \mu m^{-1}$ and $H = 1700$ Oe ($x = 1$); and d) $-D$, and e) $K_s$ vs $-D$ for $Pd_{1-x}Pt_x\ 4/\ Fe_{0.6}Co_{0.2}B_{0.2}\ 2.6$ bilayers with different $x$. The red solid curves in c) represent fits to the Lorentzian function; the dashed line in e) refers to the best linear fit.

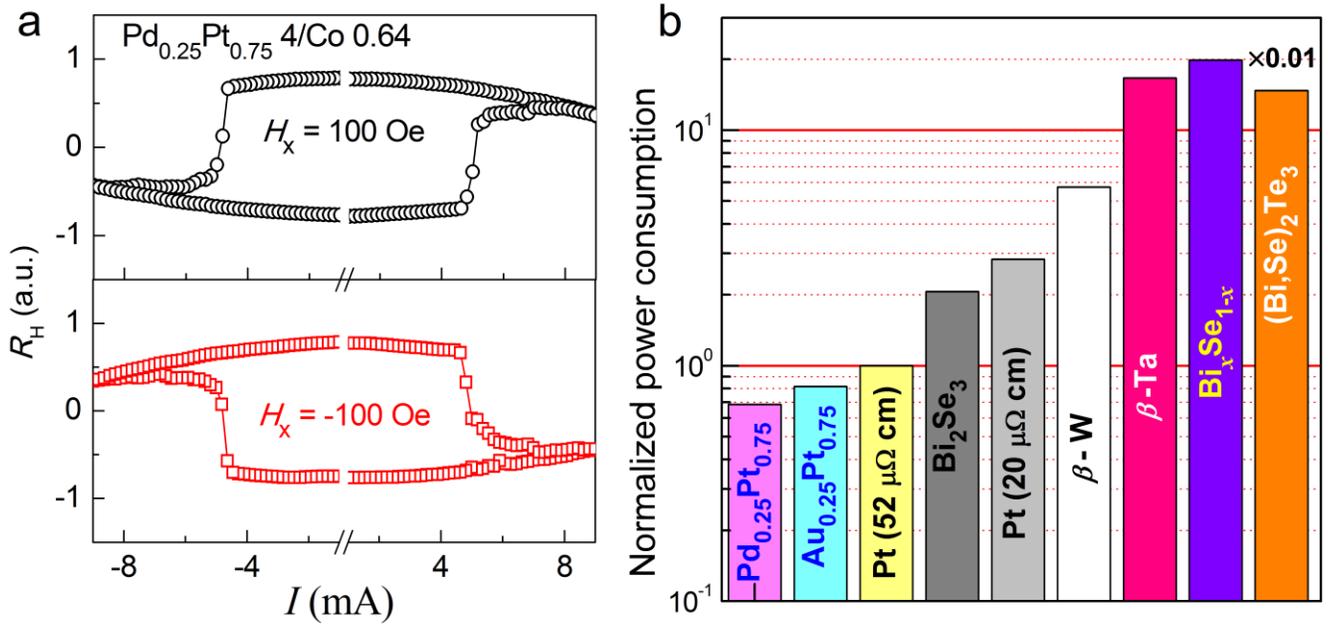

**Figure 4.** a) Deterministic current-induced magnetization switching in $Pd_{0.25}Pt_{0.75}$ 4 nm/Co 0.64 nm bilayers (effective PMA field ≈ 7.7 kOe and coercivity ≈ 0.44 kOe) with a bias field $H_x = \pm100$ Oe along current direction; b) Power consumption of normalized SOT-MRAM application for different spin Hall channel materials listed in Table I.

**Spin-orbit torque (SOT) operations** are currently energy inefficient due to a low damping-like SOT efficiency and/or the very high resistivity of the spin Hall materials. This work establishes a very effective spin current generator $Pd_{1-x}Pt_x$ that combines a notably high energy efficiency with a very strong and tunable DMI for advanced chiral spintronics and spin torque applications.

Key words: spin-orbit torque, spin Hall conductivity, spin Hall effect, Dzyaloshinskii-Moriya interaction

Authors: L. J. Zhu,* K. Sobotkiewich, X. Ma, X. Li, D. C. Ralph, R. A. Buhrman*

Titles: Strong damping-like spin-orbit torque and tunable Dzyaloshinskii-Moriya interaction generated by low-resistivity $Pd_{1-x}Pt_x$ alloys

**ToC figure**

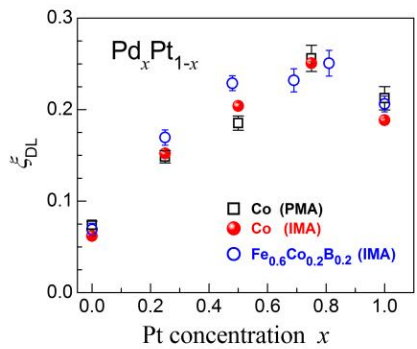

**Strong and tunable anti-damping spin-orbit torque and Dzyaloshinskii-Moriya interaction generated by low-resistivity Pd$_{1-x}$Pt$_x$ alloys**

Lijun Zhu,[1] Kemal Sobotkiewich,[2] Xin Ma,[2] Xiaoqin Li,[2] Daniel. C. Ralph,[1,3] Robert A. Buhrman[1]

1. *Cornell University, Ithaca, NY 14850, USA*
2. *Department of Physics, Center for Complex Quantum Systems, The University of Texas at Austin, Austin, Texas 78712, USA*
3. *Kavli Institute at Cornell, Ithaca, New York 14853, USA*

**A. X-Ray diffraction measurements**

**B. Out-of-plane Harmonics response measurements**

**C. In-plane harmonics response measurements**

**D. Determination of interfacial magnetic anisotropy energy density from FM thickness dependence study**

**E. Calculation of power consumption of a SOT-MRAM device**

## A. X-Ray diffraction measurements

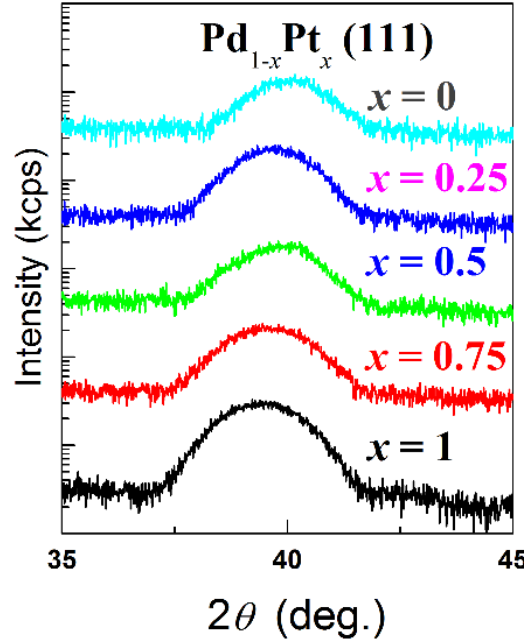

Figure S1. X-ray diffraction $\theta$-$2\theta$ patterns for Pd$_{1-x}$Pt$_x$ 4/Co $t$ ($t = 0.75$ for $x = 1$ and 0.64 for $x \leq 0.75$) with different Pd$_{1-x}$Pt$_x$ composition $x$.

## B. Out-of-plane Harmonics response measurements

For harmonic response measurements, a lock-in amplifier was used to source a sinusoidal voltage ($V_{in} = 4$ V) onto the bar (length $L = 60$ μm) orientated along the $x$ axis (see Figure S2(a)) and to detect the in-phase first and out-of-phase second harmonic Hall voltages, $V_{1\omega}$ and $V_{2\omega}$. As shown in Figure S2(b), the fairly square anomalous Hall voltage ($V_{AH}$) hysteresis loops of Pd$_{1-x}$Pt$_x$ 4/Co 0.64 bilayers and the large coercivity of 250-500 Oe evidence the strong PMA of these samples, which allows us to obtain high-quality out-of-plane harmonic response data. The damping-like ($H_{DL}$) and field-like ($H_{FL}$) effective spin-torque field can be determined following $H_{DL(FL)} = -2\frac{\partial V_{2\omega}}{\partial H_{x(y)}} / \frac{\partial^2 V_{1\omega}}{\partial^2 H_{x(y)}}$. We note that in analyzing our out-of-plane harmonics results we do not employ the so-called "planar Hall correction (PHC)" which was developed to account for any magnetoresistance effects in *PMA* HM/FM bilayers (the combination of anisotropic magnetoresistance in the FM and spin Hall magnetoresistance in the HM)[1] So far, there have little discussion in the published literature about the "PHC problem" although, as far as we know, it has been noticed by some other groups experienced in the harmonic technique. We do note that recently M. Hayashi, the developer of this "PHC", has noted the existence of the problem. [2] Experimentally, this "PHC" is not of a concern when the HM/FM systems have very poor interfacial spin transparency and very small spin torques are exerted because in that case, the changes in the measured spin torques due to this "PHC" are boringly small. The "PHC" is also found not to be important for high-spin transparency HM/FM systems when the ratio of the planar Hall voltage ($V_{PH}$) and anomalous Hall voltage ($V_{AH}$), $\xi = V_{PH}/V_{AH}$ is not larger than ~0.1 as this correction actually has negligible influence on the value of the damping-like effective spin torque field (efficiency). However when $\xi$ approaches 0.5, this "PHC" gives an infinite change, which certainly indicates the invalidity of the "PHC" in that case. In fact, when $\xi$ is larger than typically > 0.2, we find this "correction" leads to larger numbers for the dampinglike spin-torque efficiency than it should be in all cases, in some cases to unphysical (sign reversal) values, that are in sharp disagreement with those values obtained from our other technique, i.e. the harmonic response measurements on the in-plane magnetized samples. Neglecting the PHC for the PMA samples gives results that are in close accord with the results from the IMA samples (see Figure 1c in the main text for a comparison of the dampinglike spin torque efficiencies from different techniques and different HM/FM systems). The clarifying of possible origins of this absence of magnetoresistance effects in the PMA harmonic measurements is beyond the scope of this paper and will be discussed elsewhere in detail. Here we show a few examples for the absence of such "PHC" in Table S1. The "PHC", if applied, will give unreasonably larger dampinglike spin torque efficiency $\xi_{DL}$ for the PMA Pt and $Pd_{0.25}Pt_{0.75}$ samples, and unphysical sign change in case of PMA Pd samples. In all the cases, the intentive application of such "PHC" leads to fieldlike spin torque efficiency $\xi_{FL}$ that is unphysical in magnitude and sign. We also see such invalidaty (sign change) of "PHC" for $\beta$-W samples.

Table S1  Examples for the absence of the so-called "planar Hall correction".

| | $\xi = V_{PH}/V_{AH}$ | Out-of-plane harmoncis No "PHC" | | Out-of-plane harmoncis With "PHC" | | In-plane harmonics |
|---|---|---|---|---|---|---|
| | | $\xi_{DL}$ | $\xi_{FL}$ | $\xi_{DL}$ | $\xi_{FL}$ | $\xi_{DL}$ |
| Pt/Co | 0.31 | 0.21 | -0.049 | 0.30 | 0.14 | 0.19 |
| $Pd_{0.25}Pt_{0.75}$/Co | 0.28 | 0.26 | -0.048 | 0.33 | 0.13 | 0.25 |
| Pd/Co | 0.56 | 0.07 | -0.050 | -0.1 | -0.16 | 0.06 |

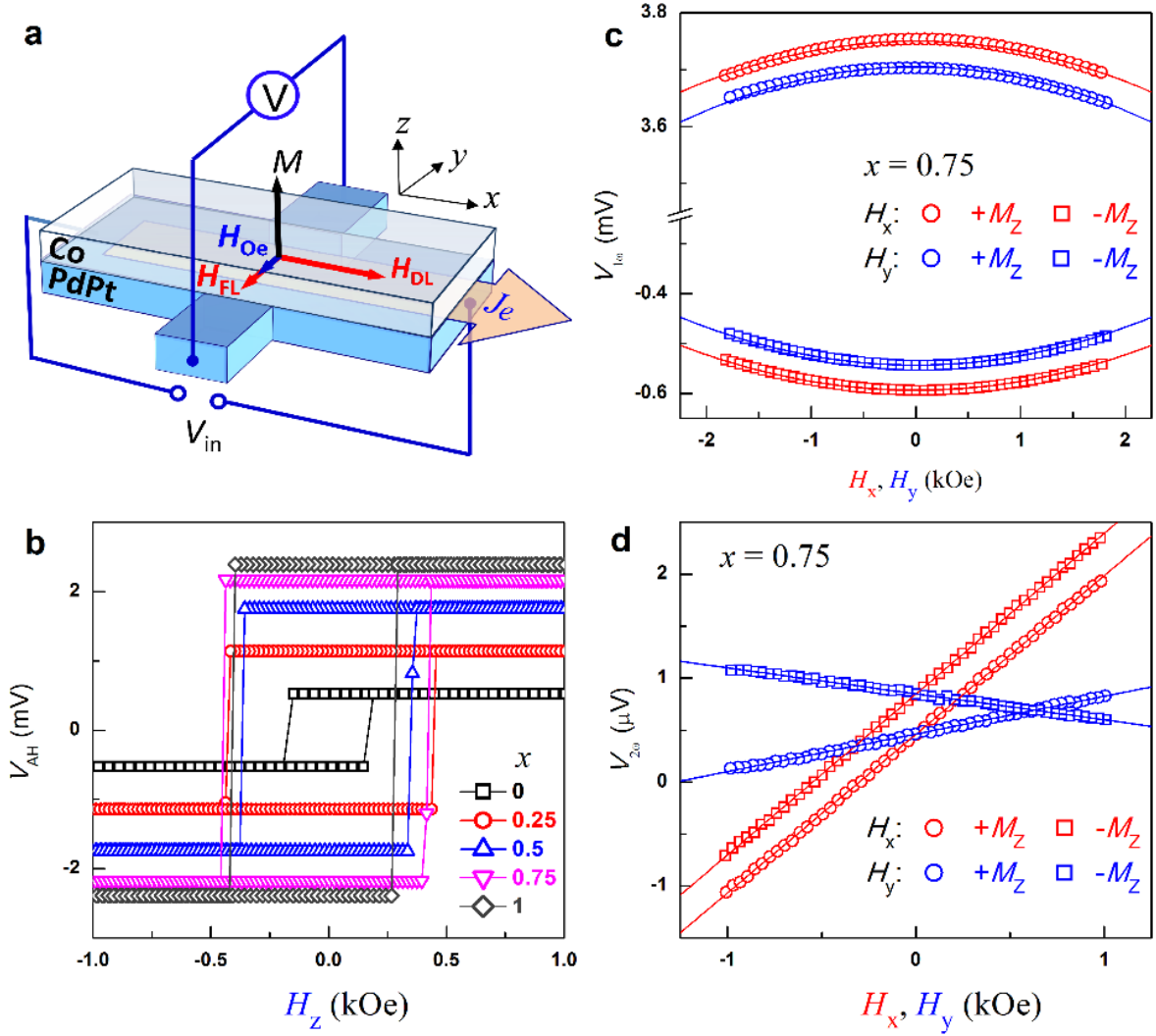

**Figure S2.** Out-of-plane harmonic response measurement. **a)** Schematic of the Hall bar devices and the measurement coordinate, a 4 V sinusoidal voltage was sourced onto the Hall bar along the $x$ axis for all the harmonic response measurements; **b)** Anomalous Hall voltage hysteresis loops for different $x$; **c.** The first ($V_{1\omega}$) and **d)** second ($V_{2\omega}$) harmonic voltages plotted as a function of in-plane fields $H_x$ (red) and $H_y$ (blue), respectively (for $x = 0.75$). For clarity, the $H_y$-dependent $V_{1\omega}$ data in **d)** are artificially shifted by ±0.001 mV for $\mp M_z$, respectively.

## C. In-plane harmonics response measurements

For in-plane magnetized HM/FM bilayer, the in-plane angle ($\varphi$) dependence of $V_{2\omega}$ is given by $V_{2\omega} = V_a\cos\varphi + V_p\cos\varphi\cos2\varphi$, where $V_a = -V_{AH}H_{DL}/2(H_{in}+H_k)+V_{ANE}$, and $V_p = -V_{PH}H_{FL}/2H_{in}$ with $V_{AH}$, $V_{ANE}$, $H_{in}$, $H_k$, $V_{PH}$, $H_{DL}$, and $H_{FL}$ being anomalous Hall voltage, anomalous Nerst effect, in-plane bias field, perpendicualr anisotropy field, planar Hall voltage, damping-like effective spin torque field, and field-like effective spin torque field. $\varphi$ dependence of $V_{1\omega}$ is given by $V_{1\omega} = V_{PH}\cos2\varphi$, from which the planar Hall voltage $V_{PH}$ can be determined. We separated the damping-like term $V_a$ and the field-like term $V_p$ for each $H_{in}$ and each $x$ by fitting $V_{2\omega}$ data to $V_{2\omega} = V_a\cos\varphi + V_p\cos\varphi\cos2\varphi$ (see Fig. S3(a)). The linear fits of $V_a$ versus $-V_{AH}/2(H_{in}+H_k)$ and $V_p$ versus $-V_{PH}/2H_{in}$ (see Fig. S3(b)) yield the values of $H_{DL}$ and $H_{FL}$ for $Pt_x(MgO)_{1-x}$/Co bilayers. Here $V_a = -H_{DL}V_{AH}/2(H_k+H_{in})+V_{ANE}$, $V_p = -V_{PH}H_{FL}/2H_{in}$ where $V_{PH}$, $V_{ANE}$, and $H_{in}$ are the planar Hall voltage, the anomalous Nernst effect voltage, and the applied in-plane magnetic field, respectively.

Here we further discuss the thermoelectronic effects. For HM/FM bilayer, there could be a vertical thermal gradient generated due to the Joule heating and asymmetric thermal dissipation into the substrate and capping layer. This vertical thermal gradient can contribute a transverse voltage that scales with $\nabla T \times M$ via either anomalous Nernst effect (ANE) and longitudinal spin Seebeck effect (SEE). Here the symmetry of the longitudinal SEE is the same as that of the ANE signal, while the longitudinal SEE signal is generally negligible compared to the ANE signal ($V_{ANE}+V_{SEE} \approx V_{ANE}$).[3] For perpendicularly magnetized bilayers, e.g. Pt/Co, the harmonic response voltages are not influenced by the thermal gradient as $\nabla T \times M = 0$. For in-plane harmonic response measurements, the signal contributions of ANE and SEE are subtracted by the linear fits of $V_a$ versus $-V_{AH}/2(H_{in}+H_k)$ and $V_p$ versus $-V_{PH}/2H_{in}$ yield the values of $H_{DL}$ and $H_{FL}$ for $Pt_x(MgO)_{1-x}$/Co bilayers.

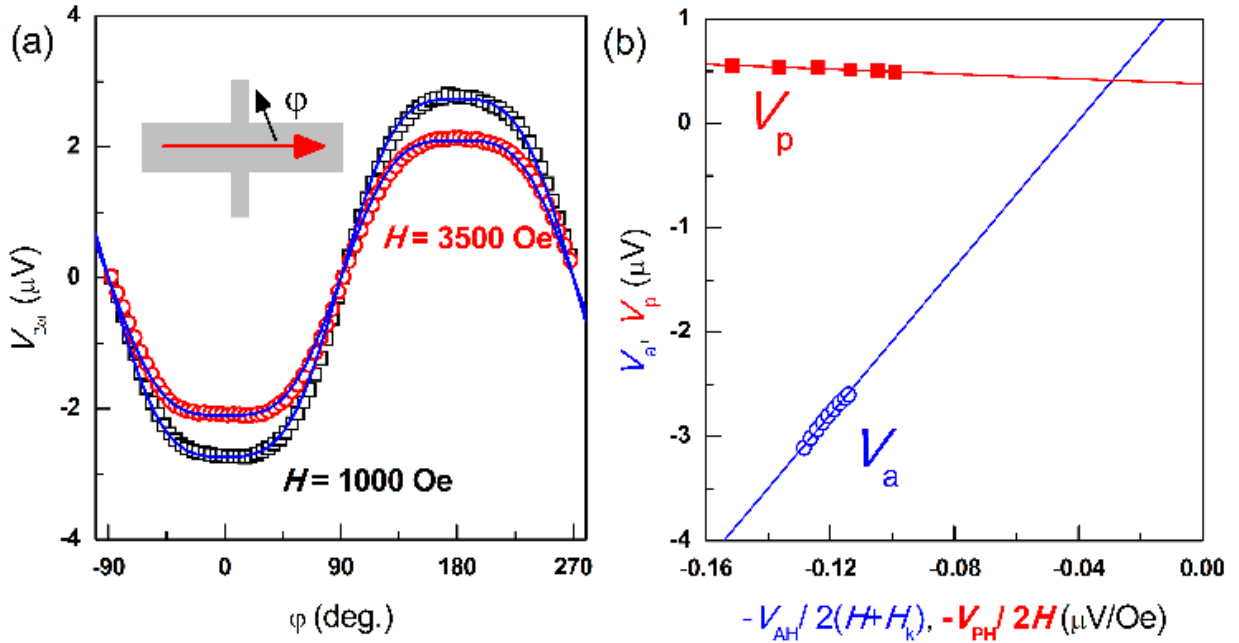

**Figure S3.** In-plane harmonic response measurement. **a)** In-plane angle ($\varphi$) dependence of $V_{2\omega}$ for the $Pd_{0.25}Pt_{0.75}$ 4/$Fe_{0.6}Co_{0.2}B_{0.2}$ 2.8 bilayer under in-plane bias fields with magnitudes of 1000 (blue) and 3500 Oe (red), the solid lines refer to the best fit to $V_{2\omega} = V_a\cos\varphi - V_p\cos2\varphi\cos\varphi$, where $V_a = -H_{DL}V_{AH}/2(H_k+H_{in})+V_{ANE}$, $V_p = -V_{PH}H_{FL}/2H_{in}$ with $V_{PH}$, $V_{ANE}$, and $H_{in}$ are the planar Hall voltage, the anomalous Nernst effect voltage, and the applied in-plane magnetic field, respectively. **b)** $V_a$ vs -$V_{AH}/2(H_k+H_{in})$ and $V_p$ vs $-V_{PH}/2H_{in}$ for the $Pd_{0.25}Pt_{0.75}$ 4/$Fe_{0.6}Co_{0.2}B_{0.2}$ 2.8 bilayer, where $V_a$ and $V_p$ determined by fitting the $\varphi$ dependence of $V_{2\omega}$. The slopes of the linear fits give $H_{DL}$ and $H_{FL}$.

## D. Determination of interfacial magnetic anisotropy energy density from FM thickness dependence study

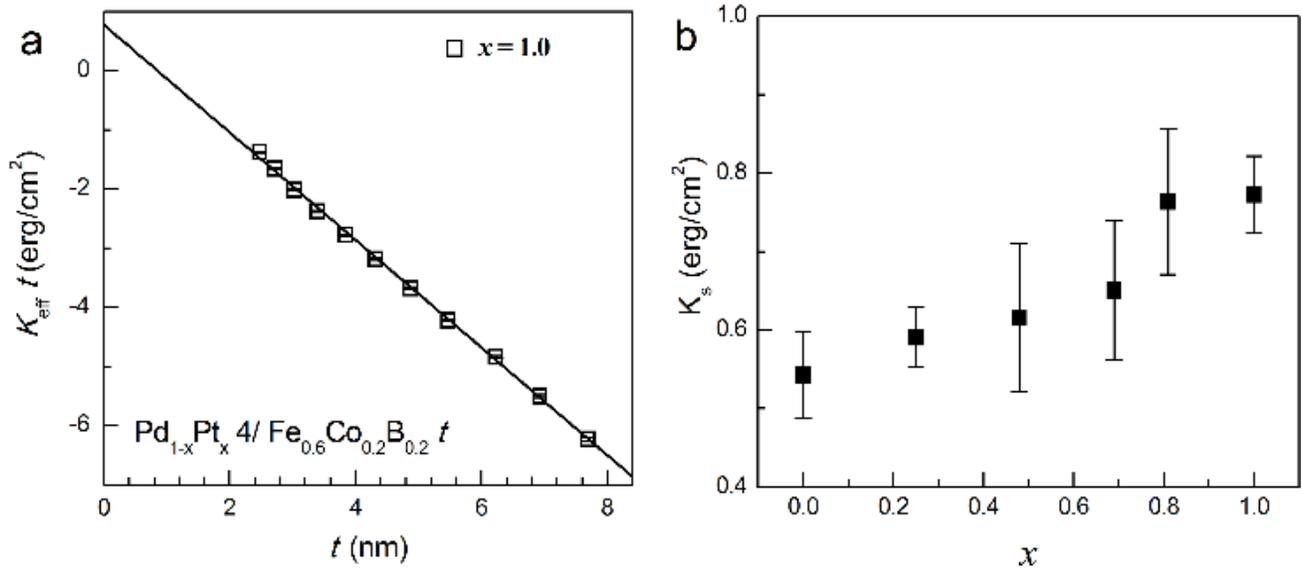

**Figure S4| Determination of interfacial magnetic anisotropy energy density. a)** $K_{eff}t$ vs $t$ for Pt 4/ Fe$_{0.6}$Co$_{0.2}$B$_{0.2}$ $t$ bilayers. Here $K_{eff}$ = -4$\pi M_{eff} M_s$/2 is perpendicular anisotropy energy, and the effective magnetization -4$\pi M_{eff}$ = -4$\pi M_s$ +2 $K_s/t$ and saturation magnetization $M_s$ are determined by spin-torque ferromagnetic resonance and VSM, respectively; **b)** The interfacial magnetic anisotropy energy density $K_s$ determined by the intercept of linear $K_{eff}t$ - $t$ fits following $K_{eff}t = 2\pi M_s^2 t + K_s$.

## E. Calculation of power consumption of a SOT-MRAM device

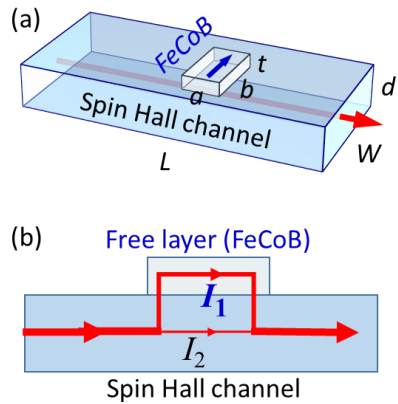

**Figure S5.** Schematics for a spin-orbit torque MRAM device. **a)** Dimension definitions. **b)** Current shunting into free layer.

Here we calculate the power efficiency of a model spin-orbit torque MRAM device consisting of a $400 \times 200 \times 4$ nm$^3$ spin Hall channel and a $110 \times 30 \times 1.8$ nm$^3$ FeCoB free layer (resistivity $\rho_{FeCoB} = 130$ $\mu\Omega$ cm) by taking into account the current shunting into the free layer. For convience of discussion, we note the device dimentions as shown in Figure S5a: channel length $L = 400$ nm, channel width $W = 200$ nm, channel thickness $d = 4$ nm; free layer "length" $a = 30$ nm, free layer "width" $b = 110$ nm, free layer thickness $t = 1.8$ nm, free layer length $a = 30$ nm, free layer width $b = 110$ nm, and free layer thickness $t = 1.8$ nm. Since $t << a$ and $t << b$, current spreading has a minimal influence and a parallel resistor model can be used. The total write current ($I_{tot}$) in the spin Hall channel is given by

$$I_{tot} = I_2 + I_1 = I_2(1 + \frac{I_1}{I_2}) = I_2(1 + \frac{\rho_{HM}}{\rho_{FM}} \frac{t\,a}{d\,W})$$  (1)

where $I_1$ is the current shunted into the FeCoB free layer and $I_2$ is the "useful" current in the spin Hall channel that drives the magnetization switching (Figure S5b). $\rho_{FM}$ and $\rho_{HM}$ are resistivity of the free layer and heavy metal layers, respectively. According to macrospin model, which is found to work reasonably well for in-plane magnetized magnetic tunneling junctions, we have

$$I_2 = \frac{2e}{\hbar} \mu_0 M_s t_{FM} \alpha (H_c + 4\pi M_{eff}\,/\,2)\,/\,\xi_{DL}$$  (2)

in which $e$, $\mu_0$, $\hbar$, $\alpha$, $H_c$, $M_{eff}$, and $\xi_{DL}$ are the elementary charge, the permeability of vacuum, the reduced Planck constant, the magnetic damping, the coercivity, the effective magnetization, the damping-like spin torque efficiency, respectively.

The power consumption is then given by:

$$P = I_1^2 \rho_{FM} \frac{a}{bt} + I_2^2 \rho_{HM} \frac{a}{Wd} + I_{tot}^2 \rho_{HM} \frac{L-a}{Wd}$$  (3)

1. M. Hayashi, J. Kim, M. Yamanouchi, and H. Ohno, Quantitative characterization of the spin-orbit torque using harmonic Hall voltage measurements, Phys. Rev. B 89, 144425 (2014).

2. Y.-C. Lau and M. Hayashi, Spin torque efficiency of Ta, W, and Pt in metallic bilayers evaluated by harmonic Hall and spin Hall magnetoresistance measurements, Jpn. J. Appl. Phys. **56**,0802B5 (2017).

3. C. O. Avci, K. Garello, M. Gabureac, A. Ghosh, A. Fuhrer, S. F. Alvarado, and P. Gambardella, Interplay of spin-orbit torque and thermoelectric effects in ferromagnet/normal-metal bilayers, Phys. Rev. B 90, 224427 (2014).